\documentclass[conference]{IEEEtran}
\ifCLASSOPTIONcompsoc
  \usepackage[nocompress]{cite}
\else
  \usepackage{cite}
\fi

\usepackage{graphicx}
\graphicspath{ {./figures/} }
\DeclareGraphicsExtensions{.eps}

\usepackage[cmex10]{amsmath}
\usepackage{booktabs}
\usepackage{multirow}

\usepackage{amsfonts}
\usepackage{amssymb}
\usepackage{bm}
\usepackage{balance}

\usepackage{footnote}
\makesavenoteenv{tabular}
\usepackage[flushleft]{threeparttable}
 
\newcommand{\R}{\mathbb{R}}

\begin{document}
%
\title{On Feature Reduction using Deep Learning\\for Trend Prediction in Finance}

\author{\IEEEauthorblockN{Luigi Troiano}
\IEEEauthorblockA{Dept. of Engineering\\
University of Sannio\\
I-82100 Benevento, Italy\\
Email: troiano@unisannio.it}
\and
\IEEEauthorblockN{Elena Mejuto}
\IEEEauthorblockA{Dept. of Engineering\\
University of Sannio\\
I-82100 Benevento, Italy\\
Email: mejutovilla@unisannio.it}
\and
\IEEEauthorblockN{Pravesh Kriplani}
\IEEEauthorblockA{CISELab\\
University of Sannio\\
I-82100 Benevento, Italy\\
Email: pravesh.kriplani@ciselab.org}}


%


\maketitle

\begin{abstract}
One of the major advantages in using Deep Learning for Finance is to embed a large collection of information into investment decisions. A way to do that is by means of compression, that lead us to consider a smaller feature space. Several studies are proving that non-linear feature reduction performed by Deep Learning tools is effective in price trend prediction. The focus has been put mainly on Restricted Boltzmann Machines (RBM) and on output obtained by them. Few attention has been payed to Auto-Encoders (AE) as an alternative means to perform a feature reduction. In this paper we investigate the application of both RBM and AE in more general terms, attempting to outline how  architectural and input space characteristics can affect the quality of prediction.
\end{abstract}


%
\IEEEpeerreviewmaketitle

\section{Introduction}

Deep Learning (DL) is disclosing new possibilities to automate complex decision making, and Finance is one the field that can benefit more from that. The need for investment decisions to look at a wider range of information has driven the interest towards the experimentation of DL in Finance, due to the capability of the new architectures to explore relationships within groups of information sources or between sources and the quality of decisions. However, the high volume and diversity of sources requires to reduce the amount of components to a set of independent/uncorrelated sources able to express the richness of available information. This belongs to the more general task of feature reduction in machine learning.

Feature reduction relies on the possibility of mapping data points from a high dimensional input space $X$ to a lower dimensional feature space $Y$, through a function $\rho: X \to Y$. The function $\rho$ can be learned from data attempting to minimize the loss of information when data points in $X$ are projected to data points in $Y$. In many circumstances, input data points are real, i.e., $X \subset \R^n$, as well as feature data points, i.e., $Y \subseteq \R^m$, with $ n \gg m$. Function $\rho$ can be linear. This is the case of PCA and other related techniques \cite{Cunningham2015}. But most effective techniques rely on a non-linear structure of $\rho$, as pointed out by Hinton and Salakhutdinov \cite{HintonSalakhutdinov2006b}. Deep Learning is providing a new class of methods that are specifically designed to perform a non-linear feature reduction. Particular attention in Finance has been paid to Restricted Boltzmann Machines (RBM) (as in \cite{Xianggao2012}). However, other possibilities are available, Auto-Encoders (AE) among them. In general, current literature in Finance focused on trend prediction performances with few or no attention to issues regarding the feature reduction step, despite the central role it plays. 

\begin{figure}[t!]
\centering
\includegraphics[width=8cm]{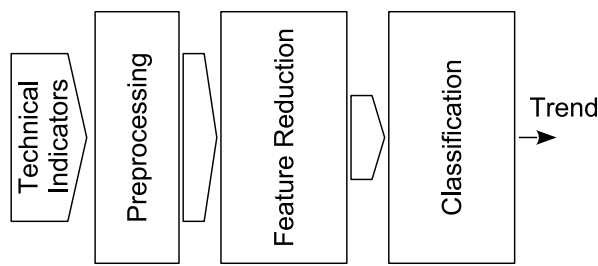}
\caption{Scheme used for for trend prediction}
\label{fig:scheme}
\end{figure}

Here, we investigate the problem looking at the different issues that can affect the quality of reduction in the problem of trend detection and prediction. Thus we considered the processing pipeline shown in Fig.\ref{fig:scheme}. As input we assume a large collection of indicators. Before we perform the feature reduction task, data are scaled. After, data are compressed and passed as input to a classifier to perform the prediction. Our interest is not regarding the performance of the classification task. For this reason we will assume only a standard SVM for prediction \cite{Cortes1995}. Instead, we are interested to better understand the impact of feature reduction on the quality of prediction, as performed by AE versus RBM, and to identify which issues regarding data source selection and preprocessing should be addressed to improve performances.

The remainder of this paper follows the following organization: Section II provides some preliminaries regarding RBM and AE; Section III discusses results of experimentation, Section IV outlines conclusions and future directions.

\section{Preliminaries}

In this sections we will provide basics of Restricted Boltzmann Machines (RBM) and Auto-Encoders (AE), that are of interest for this paper.

\subsection{Restricted Boltzmann Machine}

\begin{figure}[h!]
\centering
\includegraphics[width=6cm]{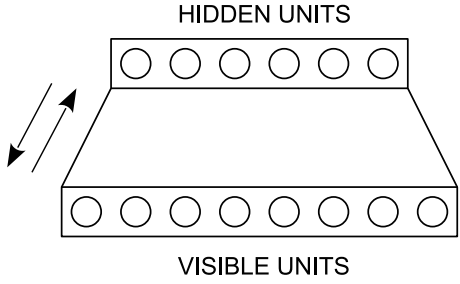}
\caption{Restricted Boltzmann Machine}
\label{fig:RBM}
\end{figure}

A Restricted Boltzmann Machine (RBM) is a network made of two layers as shown in Fig.\ref{fig:RBM}.

The input layer entails $n$ visible units $\bm V = (V_1,\ldots,V_n)$ used to represent observable data and $m$ hidden units $\bm H = (H_1,...,H_m)$ that are used to capture dependencies between observed variables. RBM have been designed to work with binary values in $\{0,1\}$. A weighting matrix $W$ defined over the relation $\bm V \times \bm H$ is used to quantify the relationship between $\bm V$ and $\bm H$. 

RBM is bidirectional. Indeed, the values given at input and hidden units are given by
\begin{equation}\label{eq:h}
h_j = \sigma \left(b_j + \sum_{i=1}^m w_{i,j} v_i \right) \qquad j = 1..m
\end{equation}
and
\begin{equation}\label{eq:v}
v_i = \sigma \left(a_i + \sum_{j=1}^n w_{i,j} h_j \right) \qquad i = 1..n
\end{equation}
where $\sigma$ is the logistic sigmoid, $a_i$ and $b_j$ the biases.

RBM belongs to the class of energy based models (EBM). Indeed, RBM can be regarded as a Markov random field with associated a bipartite undirected graph. Therefore, the values given at visible and hidden units can be interpreted in terms of conditional probabilities, that is
\begin{equation}
P(H_j=1| \bm V ) = h_j \qquad j = 1..m
\end{equation}
and
\begin{equation}
P(V_i=1| \bm H ) = v_i \qquad i = 1..n
\end{equation}

Being the RBM based on a bipartite graph, the hidden variables are mutually independent given the visible variables and vice versa. Therefore, the conditional probabilities are given as
\begin{equation}
P(\bm V| \bm H)=\prod\limits_{i=1}^{n} P(V_{i}| \bm H)
\end{equation}
\begin{equation}
P(\bm H| \bm V ) = \prod\limits_{j=1}^m P(H_j| \bm V)
\end{equation}

They can be both expressed in terms of joint probability $P(\bm V, \bm H)$ and its marginal probabilities $P(\bm V) = \sum\limits_{\bm H} P(\bm V, \bm H)$ and $P(\bm H) = \sum\limits_{\bm V} P(\bm V, \bm H)$. Since RBM makes use of the logistic sigmoid, the joint probability distribution is given by the Gibbs distribution
\begin{equation}
P(\bm V, \bm H) = \frac{1}{Z} e^{−E(\bm V, \bm H)}
\end{equation}
where $E(\bm V, \bm H)$ is named the \emph{energy function} and defined as
\begin{equation}
E(\bm V, \bm H) = -\sum_{i=1}^n a_i v_i - \sum_{j=1}^m b_j h_j -\sum_{i=1}^n \sum_{j=1}^m v_i w_{i,j} h_j
\end{equation}
where $Z$ is called \emph{partition function} and it is a normalizing constant used to assure that probability sums up to 1.

RBM can be adapted to process real-valued visible variables by scaling the input data to the unit interval, so that input values are interpreted as a-priori probabilities $p_i \in [0,1]$ that $V_i = 1$. 

RBM can be trained to replicate an input $\bm v$. Given the matrix $W$
\begin{equation}
\frac{\partial}{\partial w_{i,j}} log(P(\bm v)) = v_i h_j - v'_i h'_j
\end{equation}
where $h_j$ is obtained by Eq.\eqref{eq:h} and $v'_i$ is obtained by Eq.\eqref{eq:h}. At each step, the procedure makes use of the Gibbs sampling in order to get the vector $h'$, while $\bm v = \bm v_0$. Thus, assuming a gradient descendant rule, the update of weights is given as
\begin{equation}
\Delta w_{i,j} = \epsilon(v_i h_j - v'_i h'_j) 
\end{equation}
where $\epsilon$ is the learning rate.  In addition, biases are updated using rules $\Delta a=\epsilon (v-v')$, $\Delta b=\epsilon (h-h')$.

At the end of the training, the hidden units $\bm h$ offer a compression of visible inputs $\bm v$.

\subsection{Auto-Encoders}

\begin{figure}[h!]
\centering
\includegraphics[width=6cm]{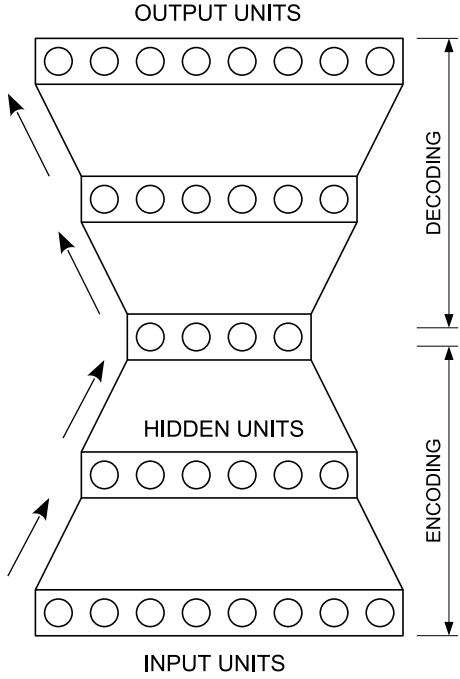}
\caption{Autoencoder}
\label{fig:AE}
\end{figure}

An Auto-Encoder (AE) is a DL network that is trained to reconstruct or approximate the input by itself. For this reason, also AEs make use of unsupervised training. AE structure consists on an input layer, an output layer and one or more hidden layers connecting them. For the purpose of reconstructing the input, the output layer has the same dimension as the input layer, forming a bottleneck structure as depicted in Fig.~\ref{fig:AE}.

An AE consists of two parts: one maps the input to a lower dimensional representation (encoding); the other maps back the latent representation into a reconstruction of the same shape as the input (decoding).
 
In the simplest structure there is just one single hidden layer. The AE takes the input $\mathbf {x} \in \mathbb {R} ^{d}={\mathcal {X}}$ and maps it onto $\mathbf {y} \in \mathbb {R} ^{p}$:
\begin{equation}
\mathbf {y} =\sigma _{1}(\mathbf {Wx} +\mathbf {b} )
\end{equation}
where $\sigma_{1}$ is an element-wise activation function such as a sigmoid function or a rectified linear unit. After that, the latent representation $\mathbf {y}$, usually referred to as code, is mapped back onto the reconstruction $\mathbf{z=x'}$ of the same shape as $\mathbf{x}$:
\begin{equation}
\mathbf {z} =\sigma _{2}(\mathbf {W'y} +\mathbf {b'} )
\end{equation}

Since we are trying to fit the model for replicating the input, the parameters ($\mathbf {W}$, $\mathbf {W'}$, $\mathbf{b}$ and $\mathbf {b}'$) are optimized so that the average reconstruction error is minimized. This error can be measured by different ways. Among them the squared error: 
\begin{equation}
{\mathcal {L}}(\mathbf {x} ,\mathbf {z} )=\|\mathbf {x} -\mathbf {z} \|^{2}=\|\mathbf {x} -\sigma _{2}(\mathbf {W'} (\sigma _{1}(\mathbf {Wx} +\mathbf {b} ))+\mathbf {b'} )\|^{2}
\end{equation}

Instead if the input is interpreted as either bit vectors, i.e., $x_i \in \{0,1\}$ or vectors of bit probabilities, i.e., $x_i \in [0,1]$, the cross-entropy of the reconstruction is a suitable solution:
\begin{equation}
{\mathcal {L}}(\mathbf {x} ,\mathbf {z} )= - \sum _{k=1}^{d} [x_k \log z_k + (1-x_k)\log (1-z_k) ]
\end{equation}

In order to force the hidden layer to extract more robust features we train the AE to reconstruct the input from a corrupted version of by discarding some of the values. This is done by setting randomly some of the inputs to zero \cite{Vincent2008}. This version of AE is called Denoising Auto-Encoder.

\section{Experimental results}

\subsection{Input Features and Data Labeling}

Historical data consist of the price series of S\&P 500 from 01 Jan 2007 to 01 Jan 2017. The input is made of multiple technical indicators computed over the price series. Table~\ref{tab:indicators} provides the list of indicators used in our experiments (a detailed description can be found in, e.g., \cite{metatrader5,amibroker,stockcharts}).
\begin{table}[h!]
  \centering
  \caption{List of indicators}
  \label{tab:indicators}
  \begin{tabular}{ll}
    \toprule
    Indicator name & Type of Indicator\\
    \midrule
    Absolute Price Oscillator (APO) & Type of Indicator\\
    Aroon &  Momentum\\
    Aroon Oscillator & Momentum \\
    MESA Adaptive Moving Average (MAMA) & Overlap studies \\
    Average Directional Movement Index (ADX) & Momentum \\
    Average Directional Movement Index Rating & Momentum \\
    Average True Range (ATR) & Volatility \\
    Balance of Power (BOP) & Momentum \\
    Bollinger Bands (BBANDS) & Overlap studies \\
    Bollinger Bandwidth & Overlap studies \\
    \%B Indicator & Overlap studies \\
    Chaikin A/D Oscillator & Volume \\
    Chande Momentum Oscillator (CMO) & Momentum \\
    Commodity Channel Index (CCI) & Momentum \\
    Directional Movement Index & Momentum \\
    Double Exponential Moving Average (DEMA) & Overlap studies \\
    Exponential Moving Average (EMA) & Overlap studies \\
    Kaufman's Adaptive Moving Average (KAMA) & Overlap studies \\
    Minimum and Maximum value over period & - \\
    Moving Average (MA) & Momentum \\
    Moving Average Convergence/Divergence (MACD) & Momentum \\
    Momentum & Momuentum\\
    Money Flow Index (MFI) & Momuentum\\
    On Balance Volume & Volume\\
    Percentage Price Oscillator (PPO) & Momuentum\\
    Plus Directional Indicator & Momuentum\\
    Plus Directional Movement & Momuentum\\
    Relative Strength Index (RSI) & Momuentum\\
    Relative Vigor Index (RVI) & Momuentum\\
    Rate of change ratio (ROC) & Momuentum\\
    Parabolic SAR & Overlap studies\\
    Stochastic Oscillator & Momentum\\
    Triple Exponential Moving Average (TEMA) & Overlap studies \\
    Triangular Moving Average (TRIMA) & Overlap studies \\
    1-day ROC of a Triple Smooth EMA (TRIX) & Momentum\\
    Ultimate Oscillator & Momuentum\\
    Weighted Moving Average (WMA) & Overlap studies \\
    Williams' Percent Range (\%W) & Momuentum\\
    \bottomrule
  \end{tabular}
\end{table}

\begin{figure}
\centering
\includegraphics[width=8cm]{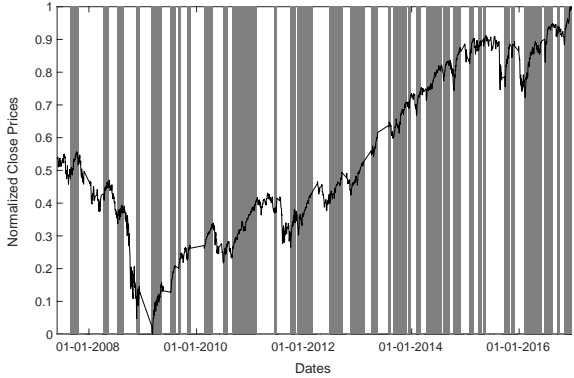}
\caption{S\&P 500 index with trend labeling (normalized values; gray bands for uptrend and white bands for downtrend).}\label{fig:labeling}
\end{figure}

Trend labeling of data is performed by assigning at each time $t$ a value $y(t) \in \{+1;-1\}$ for uptrend and downtrend respectively. The rule followed to assign the label makes use of a \textit{centered} moving average (cMA) of the index at time $t$ using the rolling window $[t-3, t+3]$. After, the following criteria for labeling are applied:
\[y(t)=+1 \;\text{if} \; \text{cMA}(t)>\text{close(t)}\; \text{and}\; \text{cMA}(t+3)>\text{cMA(t+1)}\]
\[y(t)=-1 \;\text{if} \; \text{cMA}(t)<\text{close(t)}\; \text{and}\; \text{cMA}(t+3)<\text{cMA(t+1)}\]
Otherwise at time $t$ we keep the previous label, i.e., $y(t) = y(t-1)$. In Fig.~\ref{fig:labeling} is shown the close price series and the value of labels.

In order to improve the quality of the training set, thus predictability of model, we focus on periods larger enough to outline a trend (at least $10$ days) and consistent with value movements, meaning that an uptrend should entail a positive value increment within the period, while we should have a decrement for downtrends. Periods that do not meet these characteristics are discarded from the training set.

All data available have been used to identify the structure able to compress the input data. Instead the the SVM has been trained using the 90\% of available data and tested over the remaining 10\%. Comparisons have been performed using the most recent 10\% of data for testing. In order to avoid a bias of results due to the selection of most recent period, the final comparison between RBM and AE has been performed using a 10-fold cross validation. 

\subsection{Experiment Setting}

All the experiments are carried out on a workstation equipped with an Intel Xeon Processor E5 v3 Family, $3.5$GHz x$8$, $16$GB RAM and GPU GeForce GTX 980 Ti with $6$GB RAM on board. 

The framework has been developed in Python. The implementation of AE and RBM are based on Theano \cite{theano}, while SVM is based on scikit-learn library for machine learning \cite{scikit}. All indicators are calculated by using TA-Lib, a technical analysis library \cite{talib}.

\subsection{Model Fitting}

In order to accomplish the feature reduction we make use of AE and RBM as preliminary to a SVM classifier used for prediction. A comparison between AE and RBM have been done in terms of accuracy of prediction and time required to train the network.

The training of AE is based on Backpropagation algorithm with stochastic descendant gradient for updating weights and Cross-Entropy as loss function. The weighting matrix $W$ is initialized with values uniformly sampled in the interval $[-4\sqrt{6/(n_{visible}+n_{hidden})}, +4\sqrt{6/(n_{visible}+n_{hidden})}]$ and the biases are initialized to $0$, as suggested by \cite{Glorot10} for sigmoid activation functions.

The algorithm used to train the RBM has been the gradient-based persistent contrastive divergence learning procedure\cite{Tieleman2008}. In this case the initial values of the weights are chosen from a zero-mean Gaussian with a standard deviation of $0.01$ as suggested by Hinton in \cite{Hinton12}. Hidden and visible biases are initialized to $0$. 

In both cases we divide the training set into small \textit{mini-batches} to accelerate the computation. Also we have introduced an adaptive learning rate that is exponentially decreasing by means of a constant decay-rate, in the expectation of improving both accuracy and efficiency of the training procedures. We use a form of regularization for early-stopping to avoid over-fitting issues.

\subsection{Performance Results}

Alternatives are compared by means of prediction \emph{accuracy}, that is the number of correct labels over the overall number of labels.First we consider two important aspects regarding input data: diversity and scaling.

\vspace{3mm}
\textit{Diversity.} The first question we consider is how important is to use diverse sources of information, where by "diverse" we mostly mean independent or uncorrelated. To investigate this aspect we first consider only cross-over indicators obtained by crossing different slower and faster moving averages (MA). In particular, we consider $11$ faster MAs (with all periods within the range $[5,15]$) and $11$ slower MAs (with periods uniformly distributed within the range $[20,30]$). This leads to $121$ indicators obtained by the different combinations of MAs.

\begin{figure}[h!]
\centering
\includegraphics[width=8cm,height=4.5cm]{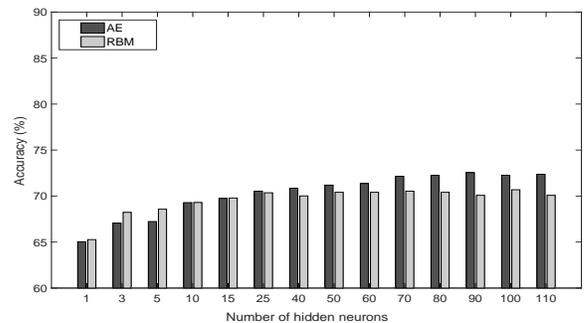}
\caption{Prediction accuracy with linearized crossovers as input features.}\label{fig:accBAR_ex1}
\end{figure}

In Table~\ref{tab:ex1} and Fig.~\ref{fig:accBAR_ex1} we report the accuracy obtained for both AE and RBM with a varying number of hidden neurons. The performance obtained of both AE and RBM are comparable, and it tends to improve by increasing the number of hidden neurons.
\begin{table}[h!]
  \centering
  \caption{Prediction accuracy with linearized crossovers as input features}
  \label{tab:ex1}
  \begin{threeparttable}
  \begin{tabular}{ccc}
    \toprule
    Number of\\ hidden neurons & AE & RBM\\
    \midrule
    1 & $65.04\%$ & $65.27\%$\\
    3 & $67.06\%$ & $68.23\%$\\
    5 & $67.23\%$ & $68.57\%$\\
    10 & $69.29\%$ & $69.3\%$\\
    15 & $69.74\%$ & $69.79\%$\\
    25 & $70.53\%$ & $70.36\%$\\
    40 & $70.86\%$ & $70.02\%$\\
    50 & $71.19\%$ & $70.41\%$\\
    60 & $71.36\%$ & $70.41\%$\\
    70 & $72.15\%$ & $70.53\%$\\
    80 & $72.26\%$ & $70.41\%$\\
    90 & $72.54\%$ & $70.08\%$\\
    100 & $72.26\%$ & $70.69\%$\\
    110 & $72.36\%$ & $70.08\%$\\
    \multicolumn{1}{c}{All \tnote{1}}& \multicolumn{2}{c}{$69.69\%$}\\
    \bottomrule
  \end{tabular}
  \begin{tablenotes}
    \item[1] In this case the indicators are provided directly to the SVM classifier
  \end{tablenotes}
  \end{threeparttable}
\end{table}

The low accuracy values obtained and the low speed in their improvement by a higher number of neurons suggest that diversity can play a relevant role. Indeed, even an high number of features may convey few information when sources are highly correlated.

In Table~\ref{tab:ex2} we report accuracy obtained by using the whole set of available indicators. Some of them are parametric with respect to the look-back period. For those we assumed different periods, namely $n=3,14,30$, in the aim of enriching the available information. We also included the adjusted closing price and volume of the index. This leads to collect a total of $93$ source, each providing a specific day-by-day feature. By looking at the results, we can observe a substantial improvement of accuracy and an initial differentiation between AE and RBM. 

\begin{table}[h!]
  \centering
  \caption{Prediction accuracy with linearized indicators as input features}
  \label{tab:ex2}
  \begin{tabular}{ccc}
    \toprule
    Number of\\ hidden neurons & AE & RBM\\
    \midrule
    1 & $61.98\%$ & $62.22\%$\\
    3 & $69.87\%$ & $71.43\%$\\
    5 & $71.54\%$ & $72.5\%$\\
    10 & $75.4\%$ & $73.62\%$\\
    15 & $76.61\%$ & $74.02\%$\\
    25 & $79.49\%$ & $73.33\%$\\
    30 & $82.03\%$ & $73.61\%$\\
    40 & $80.99\%$ & $73.56\%$\\
    50 & $81.74\%$ & $73.73\%$\\
    60 & $81.74\%$ & $73.56\%$\\
    70 & $81.92\%$ & $73.62\%$\\
    80 & $81.74\%$ & $73.21\%$\\
    90 & $81.62\%$ & $73.44\%$\\
    \multicolumn{1}{c}{All \footnote[1]{In this case the indicators are provided directly to the SVM classifier}} & \multicolumn{2}{c}{$72.40\%$}\\
    \bottomrule
  \end{tabular}
\end{table}

\begin{figure}[h!]
\centering
\includegraphics[width=8cm,height=4.5cm]{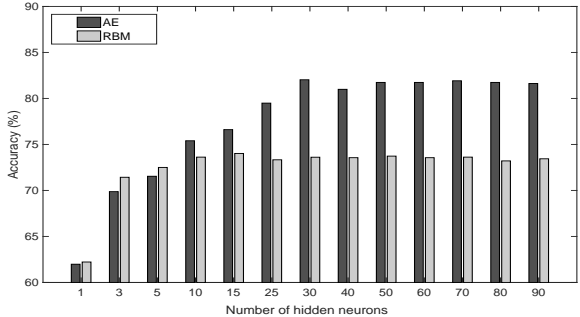}
\caption{Prediction accuracy with linearized indicators as input features.}\label{fig:accBAR_ex3}
\end{figure}

\vspace{3mm}
\textit{Scaling.} AE and RBM require to scale input values. This is generally done by a standard max/min normalization. However, other possibilities are available. Here we consider a rescaling over the unit interval $[0,1]$ obtained by means of the empirical cumulative distribution function (ECDF). The procedure consists in calculating the ECDF for each individual feature and then to assign to each instant of time $t$ its corresponding value of the ECDF. The expectation is that max/min normalization keeps unchanged the density of data points, so that information is not uniformly distributed over the unit interval. Instead, the scaling offered by ECDF is able to better distribute data points and this may contribute to improve performances. 

Table ~\ref{tab:ex3} and Fig.~\ref{fig:accBAR_ex3} show the results of this experiment. Accuracy shows an actual improvement, supporting the initial hypothesis that ECDF offers a better scaling than standard normalization.
\begin{table}[h!]
  \centering
  \caption{Prediction accuracy with indicators scaled by means of the  ECDF of their own distribution}
  \label{tab:ex3}
  \begin{tabular}{ccc}
    \toprule
    Number of\\ hidden neurons & AE & RBM\\
    \midrule
    1 & $61.98\%$ & $62.56\%$\\
    3 & $70.04\%$ & $72.11\%$\\
    5 & $73.16\%$ & $72.75\%$\\
    10 & $78.11\%$ & $75.51\%$\\
    15 & $78.4\%$ & $74.82\%$\\
    25 & $83.64\%$ & $77.02\%$\\
    30& $85.54\%$ & $76.32\%$\\
    40 & $86.18\%$ & $76.49\%$\\
    50 & $86.75\%$ & $75.92\%$\\
    60 & $84.85\%$ & $76.15\%$\\
    70 & $85.88\%$ & $75.98\%$\\
    80 & $84.97\%$ & $76.79\%$\\
    90 & $85.42\%$ & $76.05\%$\\
    \multicolumn{1}{c}{All \footnote[1]{In this case the indicators are provided directly to the SVM classifier}} & \multicolumn{2}{c}{$73.44\%$}\\
    \bottomrule
  \end{tabular}
\end{table}

\begin{figure}
\centering
\includegraphics[width=8.7cm,height=4.7cm]{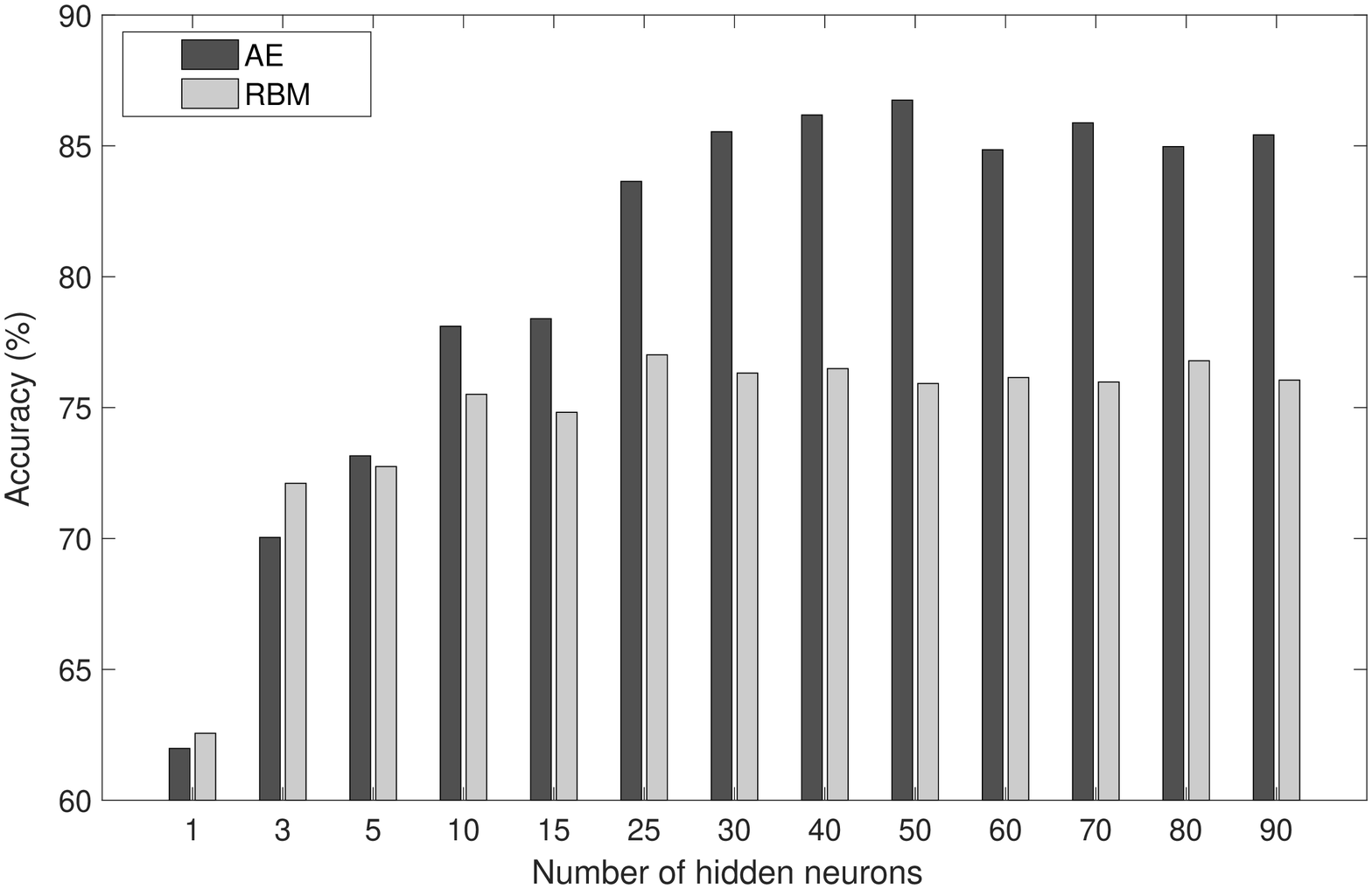}
\caption{Prediction accuracy with indicators scaled by means of the  ECDF of their own distribution.}\label{fig:accBAR_ex3}
\end{figure}

\subsection{AE Vs. RBM}

From all experiments above we can observe how there exists an optimal number of hidden neurons, over that performances do not improve or slightly decrease. That is the optimal dimensionality of the embedding performed by AE and RBM. In general, according to our experience, AE is able to reach higher dimensions. This might be the reason of better performances offered by the feature reduction based on AE. 

In order to validate this finding, we compare both the networks in their best case ($50$ hidden neurons for AE and $25$ hidden neurons for RBM, all indicators used as input, rescaled by means of ECDF) using a 10-fold cross validation procedure. In Table~\ref{tab:ex4} we report the results. They outline a consistent out-performance of AE versus RBM, resulting AE more accurate and faster to train. 

\begin{table}[h!]
  \centering
  \caption{Prediction accuracy and training time obtained with $k$-fold}
  \label{tab:ex4}
  \begin{tabular}{*{3}{p{1cm}}|*{2}{p{1cm}}}
    \toprule
     &  \multicolumn{2}{c|}{accuracy} & \multicolumn{2}{c}{training time}\\
    \midrule
    \multicolumn{1}{c}{k} & \multicolumn{1}{c}{AE} & \multicolumn{1}{c|}{RBM} & \multicolumn{1}{c}{AE} & \multicolumn{1}{c}{RBM}\\
    \midrule
    \multicolumn{1}{c}{1} & \multicolumn{1}{c}{$86.75\%$} & \multicolumn{1}{c|}{$77.02\%$} & \multicolumn{1}{c}{$16.59 \; \text{sec}$} & \multicolumn{1}{c}{$187.16 \; \text{sec}$}\\
    \multicolumn{1}{c}{2} & \multicolumn{1}{c}{$85.83\%$} & \multicolumn{1}{c|}{$76.73\%$} & \multicolumn{1}{c}{$16.63 \; \text{sec}$} & \multicolumn{1}{c}{$189.09 \; \text{sec}$}\\
    \multicolumn{1}{c}{3} & \multicolumn{1}{c}{$85.6\%$} & \multicolumn{1}{c|}{$76.09\%$} & \multicolumn{1}{c}{$15.69 \; \text{sec}$} & \multicolumn{1}{c}{$188.38 \; \text{sec}$}\\
    \multicolumn{1}{c}{4} & \multicolumn{1}{c}{$85.25\%$} & \multicolumn{1}{c|}{$75.57\%$} & \multicolumn{1}{c}{$16.27 \; \text{sec}$} & \multicolumn{1}{c}{$186.62 \; \text{sec}$}\\
    \multicolumn{1}{c}{5} & \multicolumn{1}{c}{$85.77\%$} & \multicolumn{1}{c|}{$75.63\%$} & \multicolumn{1}{c}{$15.98 \; \text{sec}$} & \multicolumn{1}{c}{$185.5 \; \text{sec}$}\\
    \multicolumn{1}{c}{6} & \multicolumn{1}{c}{$85.02\%$} & \multicolumn{1}{c|}{$75.34\%$} & \multicolumn{1}{c}{$17.16 \; \text{sec}$} & \multicolumn{1}{c}{$184.14 \; \text{sec}$}\\
    \multicolumn{1}{c}{7} & \multicolumn{1}{c}{$86.06\%$} & \multicolumn{1}{c|}{$75.51\%$} & \multicolumn{1}{c}{$17.29 \; \text{sec}$} & \multicolumn{1}{c}{$186.75 \; \text{sec}$}\\
    \multicolumn{1}{c}{8} & \multicolumn{1}{c}{$86.17\%$} & \multicolumn{1}{c|}{$76.09\%$} & \multicolumn{1}{c}{$16.16 \; \text{sec}$} & \multicolumn{1}{c}{$184.81 \; \text{sec}$}\\
    \multicolumn{1}{c}{9} & \multicolumn{1}{c}{$86.23\%$} & \multicolumn{1}{c|}{$76.04\%$} & \multicolumn{1}{c}{$17.69 \; \text{sec}$} & \multicolumn{1}{c}{$183.36 \; \text{sec}$}\\
    \multicolumn{1}{c}{10} & \multicolumn{1}{c}{$85.66\%$} & \multicolumn{1}{c|}{$76.4\%$} & \multicolumn{1}{c}{$16.25 \; \text{sec}$} & \multicolumn{1}{c}{$183.94 \; \text{sec}$}\\
    
    \bottomrule
  \end{tabular}
\end{table}

\section{Conclusions and Future Works}

Financial prediction problems often involve large data sets with complex data interactions. DL can detect and exploit these interactions that are inherently non linear and, at least currently, cannot be modeled by any existing financial economic theory. A way to do that is by identifying the underlying geometric manifold in a dimensionally reduced feature space space by means of machine learning. In this paper we investigated the application of Auto-Encoders and Restricted Boltzmann Machines able to better accomplish this task than linear methods such as PCA. The two methods have been compared in terms of trend prediction accuracy. Experiments have shown that a preliminary pre-processing of input data plays an important role. In particular values should be remapped over the unit interval $[0,1]$ taking into account their distribution of frequencies. This improves the accuracy with respect to a simple max/min normalization. In addition, diversity of input sources is crucial as well.

With respect architectures, AE performs generally better than RBM, and its training takes shorter time. Both show an optimal number of neuron, below that the feature reduction underperforms because of model underfitting, and over that value because of overfitting. The optimal cardinality of embedding neurons is larger in the case of AE, and this could explain why performances are better, as AE is able to learn a higher dimensional structure in the input data. In both architecture an adaptive learning rate is highly beneficial for improvement.

Experimental results obtained so far are preliminary and many questions are left open. Among them if using a staked AE, i.e., made of multiple hidden layers may lead to an improvement. 

\balance

\ifCLASSOPTIONcaptionsoff
  \newpage
\fi


\bibliographystyle{myIEEETran}
\bibliography{referencias.bib}

\begin{thebibliography}{10}
\providecommand{\url}[1]{#1}
\csname url@samestyle\endcsname
\providecommand{\newblock}{\relax}
\providecommand{\bibinfo}[2]{#2}
\providecommand{\BIBentrySTDinterwordspacing}{\spaceskip=0pt\relax}
\providecommand{\BIBentryALTinterwordstretchfactor}{4}
\providecommand{\BIBentryALTinterwordspacing}{\spaceskip=\fontdimen2\font plus
\BIBentryALTinterwordstretchfactor\fontdimen3\font minus
  \fontdimen4\font\relax}
\providecommand{\BIBforeignlanguage}[2]{{%
\expandafter\ifx\csname l@#1\endcsname\relax
\typeout{** WARNING: IEEEtran.bst: No hyphenation pattern has been}%
\typeout{** loaded for the language `#1'. Using the pattern for}%
\typeout{** the default language instead.}%
\else
\language=\csname l@#1\endcsname
\fi
#2}}
\providecommand{\BIBdecl}{\relax}
\BIBdecl

\bibitem{Cunningham2015}
J.~P. Cunningham and Z.~Ghahramani, ``Linear dimensionality reduction: Survey,
  insights, and generalizations,'' \emph{J. Mach. Learn. Res.}, vol.~16, no.~1,
  pp. 2859--2900, Jan. 2015.

\bibitem{HintonSalakhutdinov2006b}
G.~E. Hinton and R.~R. Salakhutdinov, ``Reducing the dimensionality of data
  with neural networks,'' \emph{Science}, vol. 313, no. 5786, pp. 504--507,
  Jul. 2006.

\bibitem{Xianggao2012}
S.~H. X.~Cai and X.~Lin, ``Feature extraction using restricted boltzmann
  machine for stock price prediction,'' in \emph{IEEE International Conference
  on Computer Science and Automation Engineering (CSAE)}, vol.~3, May 2012, pp.
  80--83.

\bibitem{Cortes1995}
C.~Cortes and V.~Vapnik, ``Support-vector networks,'' \emph{Mach. Learn.},
  vol.~20, no.~3, pp. 273--297, Sep. 1995.

\bibitem{Vincent2008}
Y.~B. Vincent~Pascal, Hugo~Larochelle and P.-A. Manzagol, ``Extracting and
  composing robust features with denoising autoencoders,'' 2008, pp.
  1096--1103.

\bibitem{metatrader5}
``Platform for forex and exchange markets,'' https://www.metatrader5.com/en.

\bibitem{amibroker}
``Exploration tool for trading,'' https://www.amibroker.com/index.html.

\bibitem{stockcharts}
``Web for financial charts,'' http://stockcharts.com/.

\bibitem{theano}
``Theano,'' http://deeplearning.net/software/theano/.

\bibitem{scikit}
``scikit-learn: Machine learning in python,'' http://scikit-learn.org/stable/.

\bibitem{talib}
``Ta-lib : Technical analysis library,'' http://ta-lib.org/.

\bibitem{Glorot10}
X.~Glorot and Y.~Bengio, ``Understanding the difficulty of training deep
  feedforward neural networks,'' in \emph{In Proceedings of the International
  Conference on Artificial Intelligence and Statistics (AISTATS)}, 2010.

\bibitem{Tieleman2008}
T.~Tieleman, ``Training restricted boltzmann machines using approximations to
  the likelihood gradient,'' in \emph{Proceedings of the 25th International
  Conference on Machine Learning (ICML)}, 2008, pp. 1064--1071.

\bibitem{Hinton12}
G.~E. Hinton, in \emph{Neural Networks: Tricks of the Trade (2nd ed.)}, ser.
  Lecture Notes in Computer Science.\hskip 1em plus 0.5em minus 0.4em\relax
  Springer, pp. 599--619.

\end{thebibliography}

\end{document}